\documentclass[12pt]{article}

\usepackage{amsmath,amssymb,amsfonts,amsbsy}
\usepackage{cite}
\usepackage{graphicx}
\usepackage{wrapfig}
\usepackage{epsfig}

\usepackage{bbm} 
\usepackage{bm} 
\usepackage{color}                                                       %
\usepackage{dsfont} 
\usepackage{latexsym} 
\usepackage{lscape} 
\usepackage{mathrsfs} 
\usepackage{morefloats} 
\usepackage{slashed} 
\usepackage{psfrag}

\def\xb{\overline{x}}

\def\vk{{\bf k}_{\perp}}

\begin{document}
\addcontentsline{toc}{subsection}{{Hard meson electroproduction and twist-3 effects.}\\
{\it S. V. Goloskokov}}

\setcounter{section}{0}
\setcounter{subsection}{0}
\setcounter{equation}{0}
\setcounter{figure}{0}
\setcounter{footnote}{0}
\setcounter{table}{0}

\begin{center}
\textbf{HARD MESON ELECTROPRODUCTION AND TWIST-3 EFFECTS.}

\vspace{5mm}

S. V. Goloskokov

\vspace{5mm}

\begin{small}
\emph{Bogoliubov Laboratory of Theoretical Physics, Joint
Institute for
Nuclear Research, Dubna 141980, Moscow region, Russia} \\
\emph{E-mail: goloskkv@theor.jinr.ru}
\end{small}
\end{center}

\vspace{0.0mm} 

\begin{abstract}
We analyze light  meson electroproduction  within the handbag
model. We study cross sections and spin asymmetries for various
mesons. The essential role of the transversity $\tilde H_T$ and
$\tilde E_T$ GPDs in electroproduction of pseudoscalar mesons is
found. Our results  are in good agrement with experiment.
\end{abstract}

\vspace{7.2mm}

In this report, investigation of the pseudoscalar meson
leptoproduction is based on the handbag approach where the leading
twist amplitude at high $Q^2$ factorizes into hard meson
electroproduction off partons and the Generalized Parton
Distributions (GPDs) \cite{sgfact}.

The amplitude of the  meson electroproduction off the proton reads
as a convolution of the partonic subprocess amplitude ${\cal H}$
and GPDs $H$
\begin{equation}\label{amptt}
{\cal M}^{a}_{\mu'\pm,\mu +} = \, \sum_{a}\,[ \langle {H}^a
  \rangle+... ] ;\;\;\langle {H}^a\rangle \propto \sum_{\lambda}
         \int_{xi}^1 d\xb
        {\cal H}^{a}_{\mu'\lambda,\mu \lambda}(Q^2,\xb,\xi)
                                   \; H^{a}(\xb,\xi,t),
\end{equation}
where  $a$ denotes the gluon and quark contribution with the
corresponding flavors;
 $\mu$ ($\mu'$) is the helicity of the photon (meson), and $\xb$
 is the momentum fraction of the
parton with helicity $\lambda$. The skewness $\xi$ is related to
Bjorken-$x$ by $\xi\simeq x/2$.

The subprocess amplitudes ${\cal H}^{V}$ are calculated within the
modified perturbative approach (MPA) \cite{sgsterman} where the
quark transverse momenta $\vk$ are taken into account together
with the gluonic radiation, condensed as a Sudakov factor. The
amplitude ${\cal H}^{V}$ contains a convolution of a
perturbatively calculated  hard part where  we keep in the
propagators the $\vk^{\,2}$ terms and the $\vk$- dependent wave
function \cite{sgkoerner}.

To estimate  GPDs, we use the double distribution (DD)
representation \cite{sgmus99}
\begin{equation} \label{ddr}
  H_i(\xb,\xi,t) =  \int_{-1}
     ^{1}\, d\beta \int_{-1+|\beta|}
     ^{1-|\beta|}\, d\alpha \delta(\beta+ \xi \, \alpha - \xb)
\, f_i(\beta,\alpha,t).
\end{equation}
The GPDs are related with PDFs through the double distribution
function
\begin{equation}\label{ddf}
f_i(\beta,\alpha,t)= h_i(\beta,t)\,
                   \frac{3}{4}
                   \,\frac{[(1-|\beta|)^2-\alpha^2]}
                          {(1-|\beta|)^{3}}\,.
                          \end{equation}
The functions $h_i$  are expressed in terms of PDFs and
parameterized as
\begin{equation}\label{pdfpar}
h(\beta,t)= N\,e^{b_0 t}\beta^{-\alpha(t)}\,(1-\beta)^{n}.
\end{equation}
Here  the $t$- dependence is considered in a Regge form and
$\alpha(t)$ is the corresponding Regge trajectory. The parameters
in (\ref{pdfpar}) are obtained from the known information about
PDFs \cite{sgCTEQ6} e.g, or from the nucleon form factor analysis
\cite{sgpauli}. The model results
 on the cross sections and spin density matrix
elements (SDME) for vector meson production obtained in
\cite{sggk05, sggk06, sggk07q, sggk08} are in good agreement with
experimental data in a wide energy range.

The hard exclusive pseudoscalar meson leptoproduction in the
leading twist is sensitive to the polarized GPDs $\tilde H$ whose
parameterization can be found in \cite{sggk07q} and $\tilde E$.
The pseudoscalar meson production amplitude with longitudinally
polarized photons ${\cal M}^{P}_{0\nu',0\nu}$ dominates at large
$Q^2$. The amplitudes with transversally polarized photons are
suppressed as $1/Q$. The pseudoscalar meson production amplitude
can be written as \cite{sggk09}:
\begin{equation}\label{pip}
{\cal M}^{P}_{0+,0+} \propto [\langle \tilde{H}^{P}\rangle
  - \frac{2\xi mQ^2}{1-\xi^2}\frac{\rho_P}{t-m_P^2}];\;
{\cal M}^{P}_{0-,0+} \propto \frac{\sqrt{-t^\prime}}{2m}\,[ \xi
\langle \widetilde{E}^{P}\rangle + 2mQ^2\frac{\rho_P}{t-m_P^2}].
\end{equation}

The first terms in (\ref{pip}) represent the handbag contribution
to the pseudoscalar (P) meson production amplitude (\ref{amptt})
calculated within the MPA with the corresponding transition GPDs.
For the $\pi^+$ production we have the $p \to n$ transition GPD
where the combination
$\tilde{F}^{(3)}=\tilde{F}^{(u)}-\tilde{F}^{(d)}$ contributes. The
second terms in (\ref{pip}) appear for charged meson production
and are connected with the P meson pole. In calculations  we use
the fully experimentally measured electromagnetic form factor of P
meson.

 In addition to the pion
pole and the handbag contribution, which in the leading twist is
determined by the $\widetilde{H}$ and $\widetilde{E}$ GPDs, a
twist-3 contribution to the amplitudes ${\cal M}_{0-,++}$ and
${\cal M}_{0+,++}$ is required to describe the polarized data at
low $Q^2$. To estimate this effect, we use a mechanism that
consists of the transversity GPD $H_T$, $\bar E_T$ in conjugation
with the twist-3 pion wave function. For the ${\cal M}_{0-,\mu+}$
amplitude we have \cite{sggk09}
\begin{equation}\label{ht}
{\cal M}^{P,twist-3}_{0-,\mu+} \propto \,
                            \int_{-1}^1 d\xb
   {\cal H}_{0-,\mu+}(\xb,...)\,[H^{P}_T+...O(\xi^2\,E^P_T)].
\end{equation}
The $H_T$ GPD is connected with transversity PDFs  as
\begin{equation}\label{et}
  H^a_T(x,0,0)= \delta^a(x);\;\;\;
\delta^a(x)=C\,N^a_T\, x^{1/2}\, (1-x)\,[q_a(x)+\Delta q_a(x)].
\end{equation}
Here we parameterize the PDF $\delta$ using the model
\cite{sgans}. The DD form (\ref{ddr},\ref{ddf}) is used to
calculate  GPD $H_T$. It is important that the $H_T^u$  and
$H_T^d$ GPDs are different in  sign.

 The twist-3 contribution to the amplitude ${\cal M}_{0+,\mu+}$
 has a form \cite{sggk11} similar to (\ref{ht})
\begin{equation}
{\cal M}^{P,twist-3}_{0+,\mu+} \propto \, \frac{\sqrt{-t'}}{4 m}\,
                            \int_{-1}^1 d\xb
 {\cal H}_{0-,\mu+}(\xb,...)\; \bar E^{P}_T.
\end{equation}
The information on $\bar E_T$ was obtained only in the lattice QCD
\cite{sglat}. The lower moments of $\bar E_T^u$  and $\bar E_T^d$
were found to be quite large, have the same sign and a similar
size.   This means that we have an essential compensation of the
$\bar E_T$ contribution in the $\pi^+$ amplitude: $\bar
E_T^{(3)}=\bar E_T^u-\bar E_T^d$. $H_T$ does not compensate in
this process. For the $\pi^0$ production we have the opposite
case. We find here a large contribution from $\bar E_T^{\pi^0}=2/3
\bar E_T^u+ 1/3\bar E_T^d$, $H_T$ effects are not so essential
here. The parameters for individual PDFs were taken from the
lattice results, and DD model was used to estimate $E_T$.

In Fig. 1a, we show the full  unseparated cross section of the
$\pi^+$ production which describes fine the HERMES data
\cite{sghermesds}.  The longitudinal cross section determined by
leading-twist dominates at small momentum transfer $-t < 0.2
\mbox{GeV}^2$. At larger $-t$ we find a not small contribution
from the transverse cross section. Effects of   $E_T$  is
negligible here.

\begin{figure}[h!]
\begin{center}
\begin{tabular}{cc}
\includegraphics[width=7.6cm,height=6.6cm]{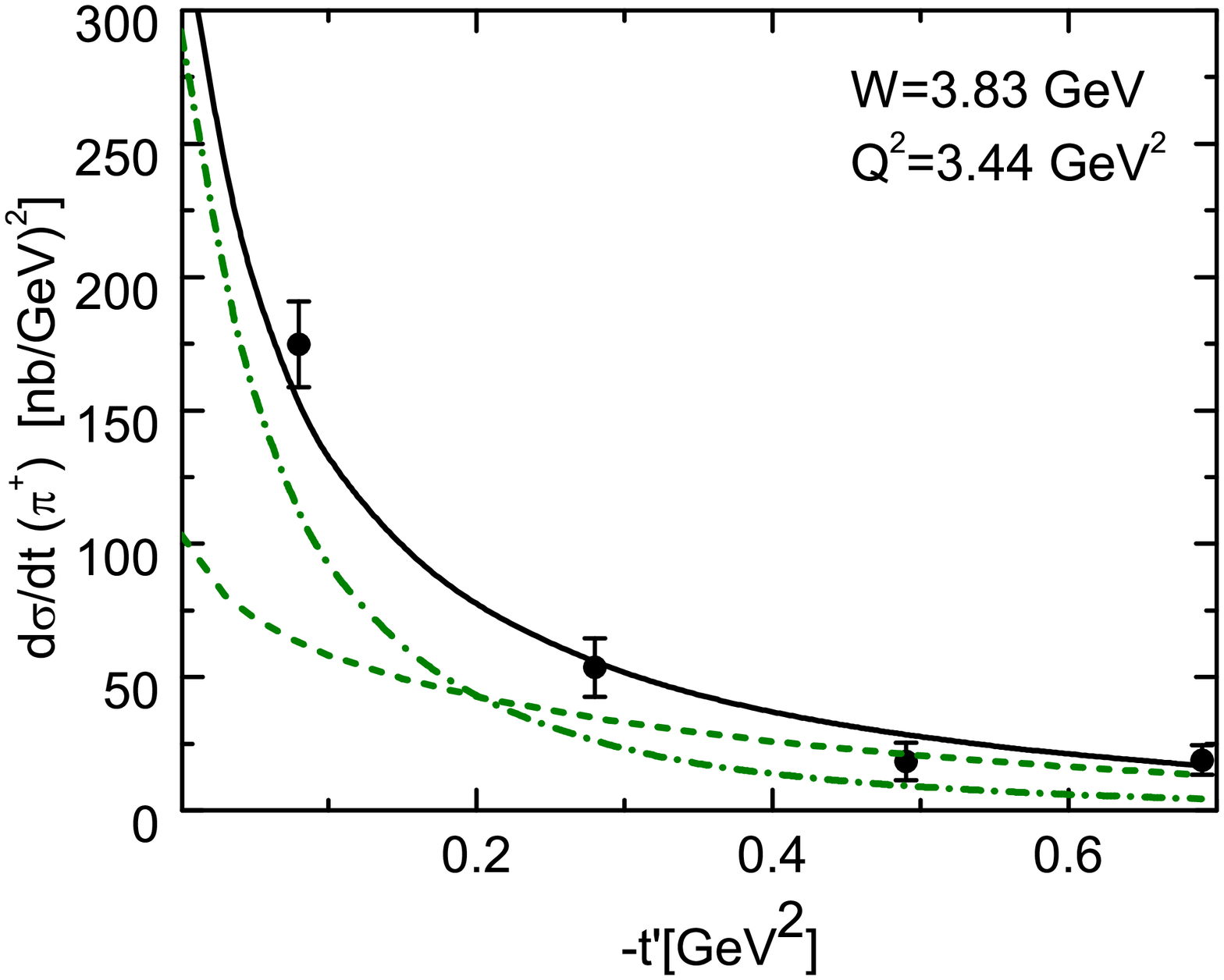}&
\includegraphics[width=7.6cm,height=6.6cm]{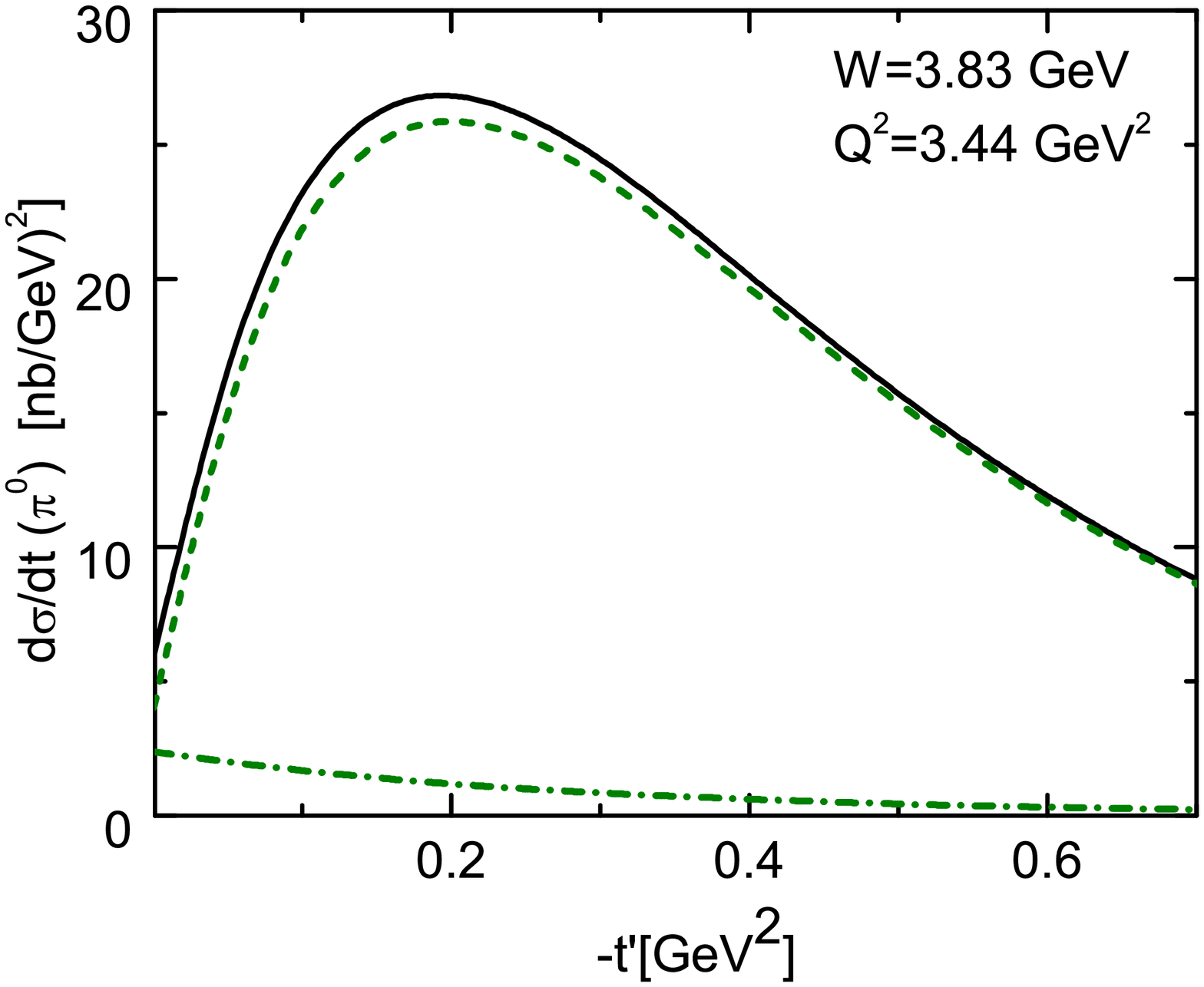}\\
{\bf(a)}& {\bf(b)}
\end{tabular}
\end{center}
\caption{{\bf(a)} The  cross section of the $\pi^+$ production
together with HERMES data. {\bf(b)} $\pi^0$ production at HERMES.
For both: full line- unseparated cross section, dashed-dotted-
$\sigma_L$, dotted line- $\sigma_T$.} \label{fig:1}
\end{figure}

\begin{figure}[h!]
\begin{center}
\begin{tabular}{cc}
\includegraphics[width=7.6cm,height=6.6cm]{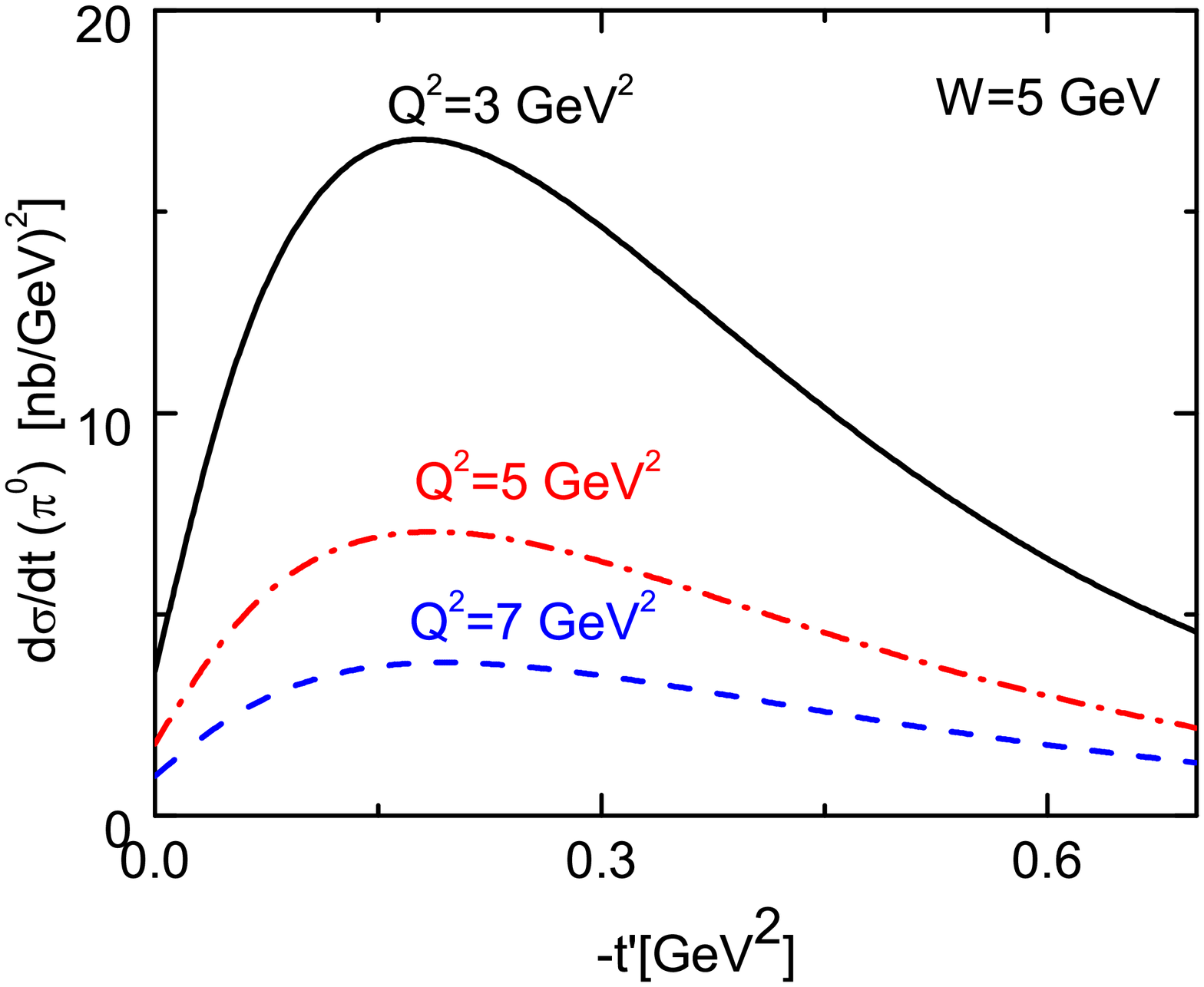}&
\includegraphics[width=7.6cm,height=6.6cm]{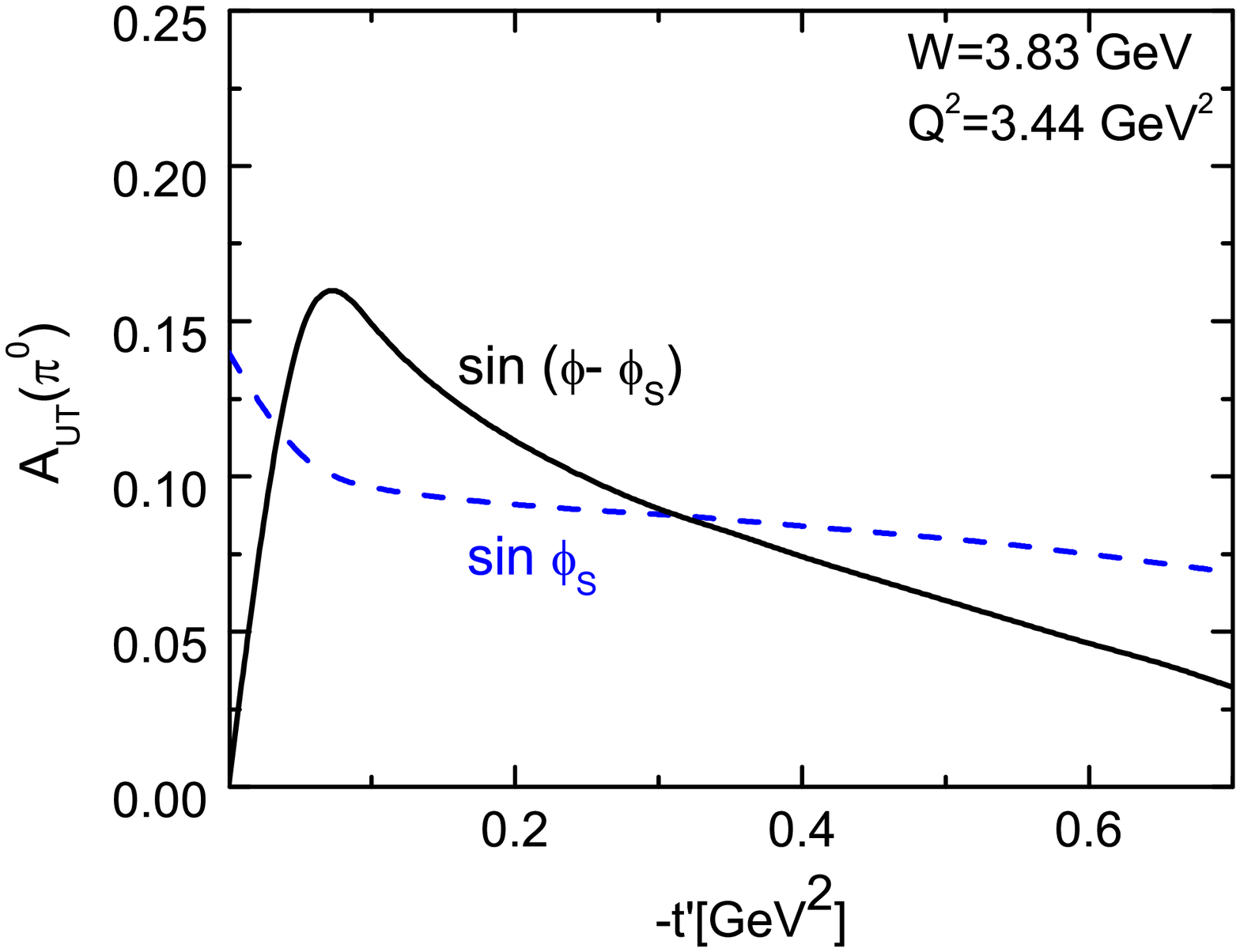}\\
{\bf(a)}& {\bf(b)}
\end{tabular}
\label{fig:2}
\end{center}
\caption{{\bf(a)} $Q^2$ dependence of $\pi^0$ production cross
section at HERMES. {\bf(b)} Predictions for the moments of
$A_{UT}$ asymmetry  of  $\pi^0$ production at HERMES.}
\end{figure}

For the $\pi^0$ production we show above that the transversity
effect should be essential. They lead to a large transverse cross
section $\sigma_T$. The longitudinal cross section, which is under
control of the leading twist contribution and expected to play an
important role, is much smaller with respect to the transverse
$\sigma_T$ cross section. The predominated role of transversally
polarized photons  is mainly generated by the $E_T$ GPDs
contribution.

This surprising result for the cross section of the $\pi^0$
production at HERMES energies \cite{sggk11} is presented in
Fig.~1b.
 It was found that the
transversity GPDs leads to a large $\sigma_T$ for all reactions of
pseudoscalar meson production with the exception of $\pi^+$ and
$\eta'$ channels \cite{sggk11}. These twist-3 effects have $1/Q$
suppression with respect to the leading twist contribution. The
$Q^2$ dependence of the transverse cross section in Fig. 2.a shows
a rapid decrease of $\sigma_T$ at  HERMES energies. It is
important that the ${\cal M}^{P,twist-3}_{0+,\mu+}$ amplitude
(\ref{et}) which is under control of $E_T$ GPDs has a zero for
$-t'=0$. This provides a minimum of the cross section at zero
momentum transfer, Figs. 1b, 2a.

In Fig. 2b, our predictions for the $\sin(\phi-\phi_s)$ and
$\sin(\phi)$ moments of $A_{UT}$ asymmetry for the transversally
polarized target are presented. Predicted asymmetries are quite
large and can be measured experimentally.

If Fig. 3a we show the ratio of the $\eta/\pi^0$ cross section at
CLAS energies for two parameterizations of $H_T$ GPDs. Different
combinations of the quark contributions to these processes leas to
the essential role of $H_T$ effects at $-t <0.2 \mbox{GeV}^2$ in
this ratio. At larger momentum transfer the $E_T$ contributions
predominate. That leads to the rapid $t$- dependence of the
$\eta/\pi^0$ cross section ratio. The preliminary CLAS data
\cite{sgclask} confirm the large $E_T$ effects in $\pi^0$
production found in the model.

\begin{figure}[h!]
\begin{center}
\begin{tabular}{cc}
\includegraphics[width=7.6cm,height=6.6cm]{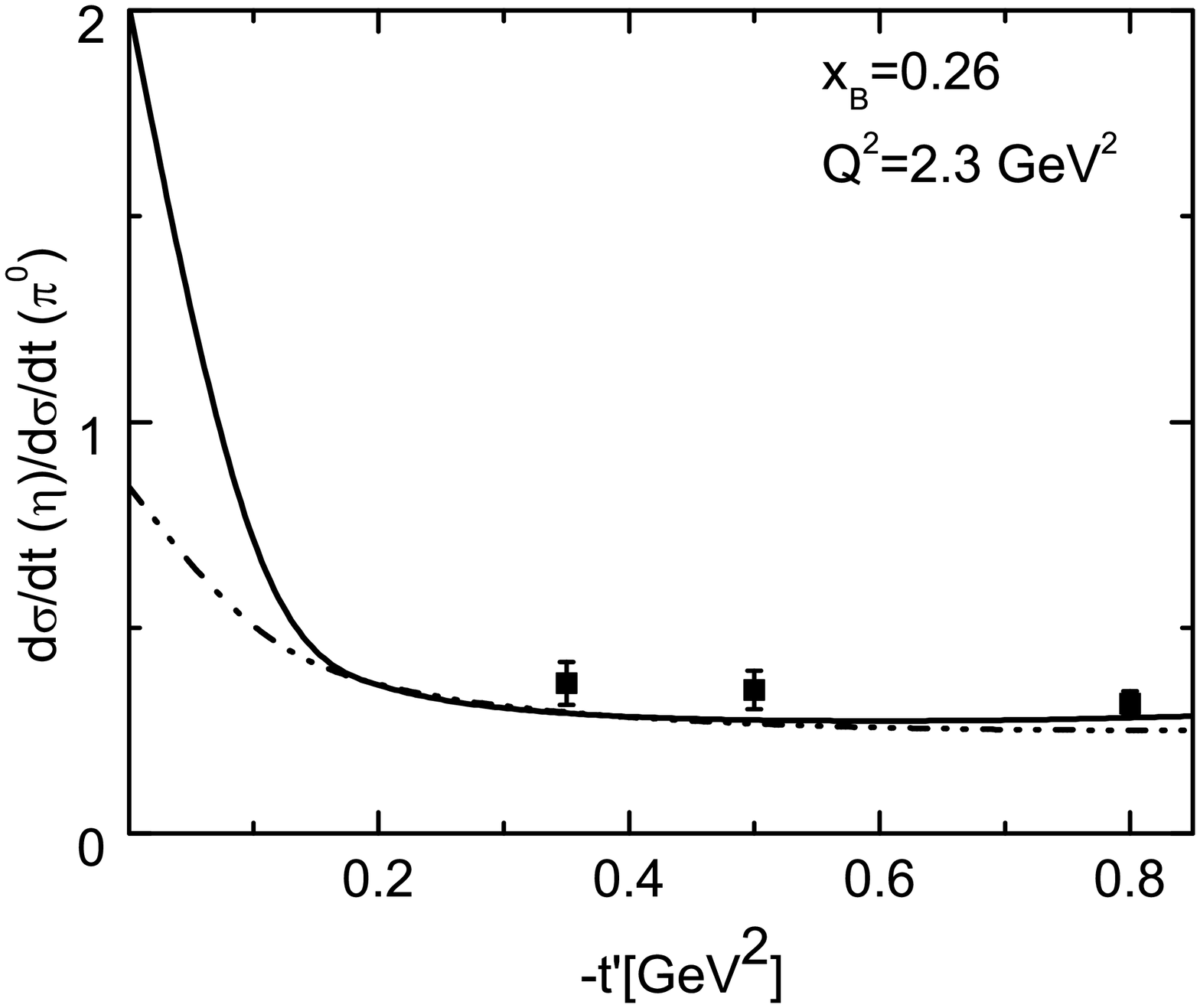}&
\includegraphics[width=7.6cm,height=6.6cm]{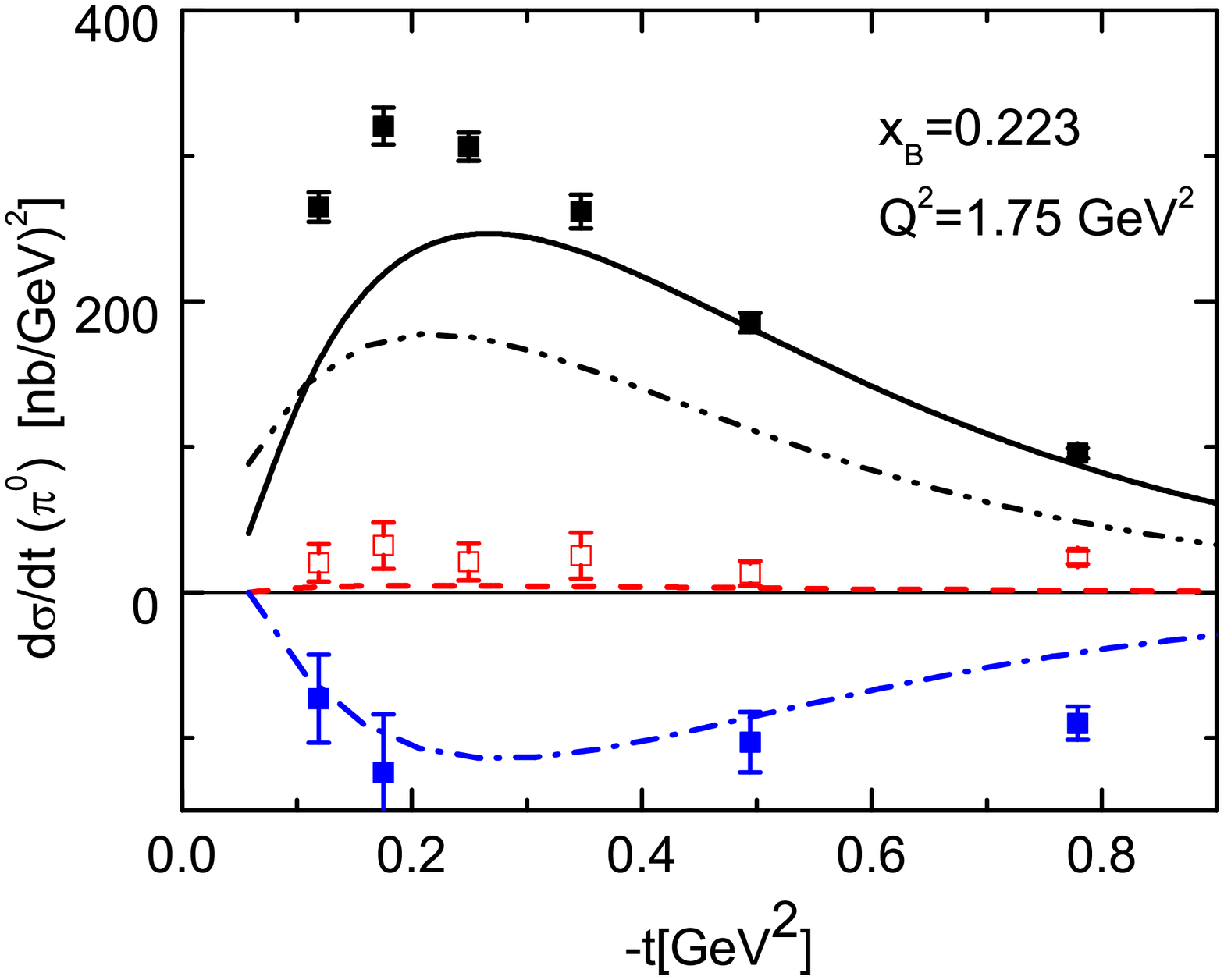}\\
{\bf(a)}& {\bf(b)}
\end{tabular}
\label{fig:3}
\end{center}
\caption{{\bf(a)} The ratio of the $\eta/\pi^0$ cross section at
CLAS together with preliminary CLAS data. {\bf(b)} The   $\pi^0$
cross section at CLAS together with preliminary CLAS data. Full
line- unseparated cross section, dashes- $\sigma_{LT}$, dashed
dotted- $\sigma_{TT}$. Dashed-dot-dotted line-- the alternative
parameterizations of $H_T$}
\end{figure}
In Fig 3b, we show our prediction for $\pi^0$ production at the
CLAS energy range together with preliminary experimental data. The
data are not far from our predictions at the CLAS energy
\cite{sgclask} and definitely show the dip at low momentum
transfer which is less with respect to the standard $H_T$
parameterization (full line). The alternative $H_T$
parameterization \cite{sggk11} shows a smaller dip at $t'=0$ and a
smaller cross section at large $t'$ as well. The main prediction
of the model- large  $\sigma_{T}$ cross section can be checked if
the data on the separated $\sigma_L$ and $\sigma_T$ cross section
will be available.

In a similar way we can estimate $E_T$ effects in the vector meson
leptoproduction. Some details can be found in \cite{sggk09}. The
$M_{0+,++}$ amplitude and correspondingly the transversity twist-3
effects are essential in the $r^1_{00}$ and $r^5_{00}$ SDME.  Our
results are shown in Fig. 4. They are consistent in signs and
values with  HERMES data \cite{sghermessd} without any free
parameters. However, such estimations now can be made only for the
quark contribution and cannot be used for the low $x_B$ range.

In this report, the  hard  pseudoscalar meson electroproduction is
calculated within the MPA   which takes into account the quark
transverse degrees of freedom and the  Sudakov suppressions. At
the leading-twist accuracy this class of reactions is sensitive to
the GPDs $\widetilde{H}$ and $\widetilde{E}$. However,  rather
strong contributions from the amplitudes $M_{0-,++}$ and
$M_{0+,++}$ are required to describe experimental data. These
amplitudes are generated by the transversity GPDs $H_T$ and $\bar
E_T$ accompanied by the twist-3 pseudoscalar meson wave functions.
Our parameterizations of  GPDs are consistent with the lattice QCD
results and other  information like nucleon form factors.
\begin{wrapfigure}[16]{R}{80mm}
  \centering 
  \vspace*{-2mm} 
  \includegraphics[width=79mm]{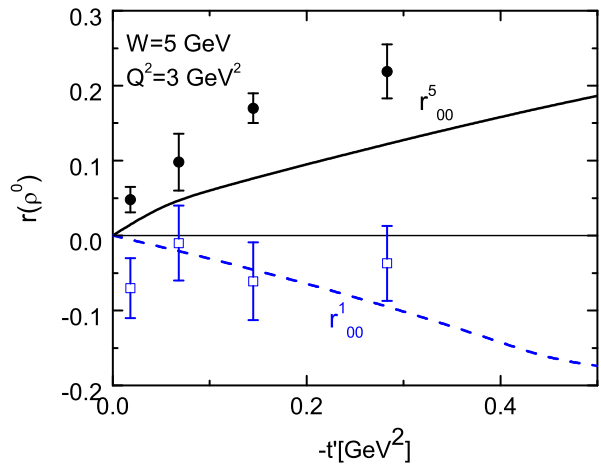}
  \caption{\footnotesize
 Twist-3 effects in spin density
matrix elements of $\rho^0$ production at HERMES.}
  \label{yourname_fig4}
\end{wrapfigure}
 The model
predicts the  large $\eta/\pi^0$ cross section ratio $\sim 1$ at
small momentum transfer and its small value $\sim .3$ at $-t'
>0.2 \mbox{GeV}^2$.  The small value of the ratio is compatible with the
CLAS data. At the same time,  JLAB data on unseparated cross
section have definite dip at $t' \sim 0$.  These model results are
determined by the  twist-3 transversity $E_T$ effects compatible
with the data.

Our calculations of the twist-3 transversity effects in SDME of
$\rho^0$ production are not far from the HERMES data.  Since our
parameterization of $\bar E_T$ fully depends on the lattice QCD
estimations, our results for the cross sections of
electroproduction of pseudoscalar mesons are real predictions. All
these observations can indicate the large transversity effects in
the mentioned reactions. To check them,  additional investigation
is needed. For example, the analysis of separated $\sigma_L$ and
$\sigma_T$ cross section in $\pi^0$ production is important to get
the definite conclusion about $\bar E_T$ GPDs.

We describe fine the well-known data on the cross section and spin
observables for various meson productions \cite{sggk05, sggk06,
sggk07q, sggk08}. We give predictions for cross sections and spin
asymmetries for all pseudoscalar meson channels \cite{sggk09,
sggk11} at low skewness and small momentum transfer. Our
predictions can be examined in future experiments and shed light
on the role of transversity effects in these reactions.

 Thus, we can conclude that
information about twist-3 transversity effects can be obtained
from pseudoscalar meson electroproduction for example at JLAB
energies.
\bigskip

This work is supported  in part by the Russian Foundation for
Basic Research, Grant  12-02-00613  and by the Heisenberg-Landau
program.
 \bigskip
\nocite{*}
\bibliographystyle{elsarticle-num}
\bibliography{martin}
 
\end{document}